\documentclass[aps, pra, showpacs, amsmath, twocolumn, groupedaddress, superscriptaddress]{revtex4}
\usepackage{graphics}
\usepackage{graphicx}
\usepackage{subfigure}
\usepackage{float}
\usepackage{longtable}
\usepackage{amssymb}
\usepackage{bm}

\begin{document}
\title{Entanglement between remote continuous variable quantum systems: effects of transmission loss}
\author{Lars Bojer Madsen}
\affiliation{Department of Physics and Astronomy,
  University of  Aarhus, 8000 {\AA}rhus C, Denmark}
\author{Klaus M{\o}lmer}
\affiliation{Department of Physics and Astronomy,
  University of  Aarhus, 8000 {\AA}rhus C, Denmark}
\affiliation{QUANTOP - Danish National Research Foundation Center for
  Quantum Optics, Department of Physics and Astronomy, University of Aarhus,
  8000 {\AA}rhus C, Denmark}

\date{\today}

\begin{abstract}

We study the effects of losses on the entanglement created between
two separate atomic gases by optical probing and homodyne detection
of the transmitted light. The system is well-described in the
Gaussian state formulation. Analytical results quantifying the
degree of entanglement between the two gases are derived and
compared with the entanglement in a pair of light pulses generated
by an EPR source. For low (high) transmission losses the highest
degree of entanglement is obtained by probing with squeezed
(antisqueezed) light. In an asymmetric setup where light is only
sent one way through the atomic samples, we find that the
logarithmic negativity of entanglement attains a constant value
$-\log_2(N)$ with $N=1/3$ irrespectively of the loss along the
transmission line.
\end{abstract}


\maketitle

\section{Introduction}
\label{Sec:Intro}

In parallel with research on implementation and manipulation of
qubits in physical quantum systems, protocols for quantum
communication and storage have been investigated, where the quantum
information is encoded in the continuous quadrature variables of the
electromagnetic field or in the collective spin of macroscopic
atomic ensembles. Successful teleportation \cite{Furusawa98},
entanglement \cite{Julsgaard01} and memory storage
\cite{Julsgaard04} have been demonstrated making use of only
Gaussian input states (coherent and squeezed states) and Gaussian
operations (homodyne detection and bilinear interactions in the
canonical conjugate $x$ and $p$ variables of the systems). The
relative simplicity of the implementation and theoretical analysis
of Gaussian states and operations comes at a prize
\cite{Eisert02,Fiurasek02,EisertPlenio}: distillation is provably
not possible! In order to distil the entanglement of Gaussian
states, one must carry out non-Gaussian operations and, at least for
a while, leave the set of Gaussian states \cite{EisertPlenio}.
Although distillation of quantum entanglement is not possible with
Gaussian states and operations, distillation of a quantum key for
cryptography has been proposed \cite{key}. Given the current
experimental interest and the relative simplicity, it is worth while
to investigate how well one may use Gaussian states and operations
to entangle remote continuous variable quantum systems, coupled by a
lossy quantum channel, and to address the related question about how
well one may teleport an unknown quantum state by use of the
entangled channel.

\begin{figure}[ttb]
  \includegraphics[width=\columnwidth]{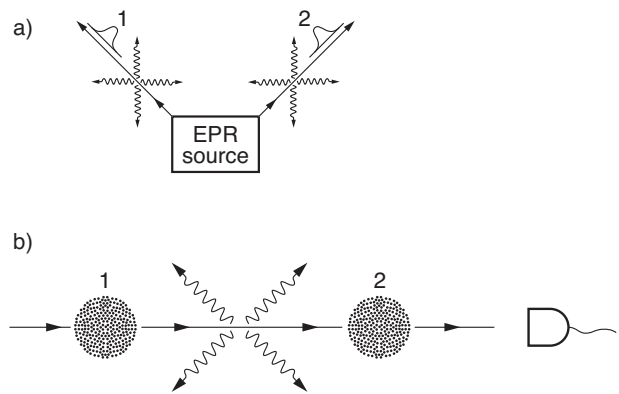}\\
\caption{Illustration of continuous variable systems sharing
entanglement between two light pulses (panel (a)) (see, e.g.,
\cite{Furusawa98}), and between two macroscopic atomic gases (panel
(b)) (see, e.g., \cite{Julsgaard01}). The losses in the transmision
lines carrying the entanglement to large distances is indicated by
the wiggly curves emerging from the transmission
lines.}\label{fig:fig1}
\end{figure}

In theory~\cite{duan00} and experiment~\cite{Julsgaard01},
entanglement between two trapped atomic gases is generated when the
atoms are probed by off-resonant light (effective Faraday rotation
interaction), and the transmitted probe light is monitored by
homodyne detection. In Fig.~\ref{fig:fig1}(b) we display the
situation we are concerned with.  We imagine the gases to be at a
large distance $L$ from each other and that there is a loss of
$\epsilon$ in the light intensity between the two gases.
Figure~\ref{fig:fig1}(a) illustrates the alternative situation where
the entanglement is shared between two light beams propagating from
an EPR source~\cite{Braunstein98,Furusawa98,Kim01,Chizhov02}. In the
present work we are primarily concerned with the system in panel
(b), and the setup in (a) with entangled photons serves as a
reference.

It is an interesting observation, that the no-distillation theorem
precludes us from establishing several such entangled ensemble pairs
and concentrating the entanglement into a single system, whereas
during the optical probing of the atomic samples, the atomic
entanglement \textit{is} indeed increasing  due to the sequential
transmission of more and more segments of the light beam, which are
also continuous variable quantum systems. It is accordingly not
clear if the entanglement achievable in a single pair of atomic
ensembles is fundamentally limited by loss and if so, what is its
maximum value. We shall answer this question by direct computation
of the state resulting from various measurement schemes.

The paper is organized as follows. In Sec.~\ref{sec:ent}, we
introduce the measure of entanglement to be used throughout this
work. In Sec.~\ref{Sec:Theory}, we recall the Gaussian covariance
matrix analysis of the continuous variable systems. In
Sec.~\ref{sec:asymmetric} we consider the atomic system
(Fig.~\ref{fig:fig1}(b)) in the asymmetric case where light is sent
in the direction from gas '1' to gas '2', only. In
Sec.~\ref{sec:symmetric} we consider a symmetrized setup with light
being sent in both directions. In Sec.~\ref{sec:quantifying} we
quantify the degree of entanglement that can be obtained in the two
systems. In Sec.~\ref{sec:Tele}, we conclude with a discussion of
the usefulness of the generated entanglement for teleportation with
continuous quantum variables.

\section{Measures of entanglement}
\label{sec:ent}
We are concerned with Gaussian continuous quantum
variables systems. A Gaussian state for a vector of variables ${\bm
y} = (y_1,y_2,\dots,y_n)$ is fully characterized by its mean value
vector $\bm m$ and its covariance matrix $\bm \gamma$ with entrance
$ij$ given by $\gamma_{ij} = 2 \text{Re} \langle (y_i-\langle
y_i\rangle)(y_j - \langle y_j \rangle)$. The degree of entanglement
of a bipartite Gaussian state is therefore also characterized by its
covariance matrix.

In our calculations below we are concerned with a two-mode
covariance matrix which in a local ($x_1,p_1,x_2,p_2)$ basis of
canonical conjugate variables reads
\begin{eqnarray}
\label{eq:cov2-tel} \gamma_{12} = \left(
  \begin{array}{cccc}
    v_{x_1} & 0 & c_x & 0 \\
    0 & v_{p_1} & 0 & c_p \\
    c_x & 0 & v_{x_2} & 0 \\
    0 & c_p & 0 & v_{p_2} \\
  \end{array}
\right),
\end{eqnarray}
with $v_{x_j}=2\text{Var}(x_j)$, $v_{p_j}=2 \text{Var}(p_j)$, and
correlations $c_\alpha$, $(\alpha=x,p)$. To quantify the
entanglement of such a state, we consider the logarithmic
negativity~\cite{Vidal02,Audenaert02}, $- \sum_{k=1}^4 \log_2
[\text{min}(1,| \lambda_k|)]$, which is calculated from the
eigenvalues $\lambda_k$ of $i \sigma^{-1} \gamma_{12}^{T_1}$, where
$\gamma_{12}^{T_1}$ is the partial transpose of $\gamma_{12}$
obtained by multiplying all entrances in $\gamma_{12}$ coupling
$p_1$ and other observables by $-1$ (i.e., $c_p \rightarrow - c_p$
in \eqref{eq:cov2-tel}), and where $\sigma_{jk}=-i [y_j, y_k]$ is
the matrix specifying the commutators between our $x$ and $p$
variables. Here and throughout, $\hbar=1$. For a matrix of the above
form, we find $| \lambda_1| = | \lambda_2|$ and $|\lambda_3| =
|\lambda_4|$ and only one, say $|\lambda|$ of these norms may be
smaller than unity and hence contribute to the measure of
entanglement. We label this quantity $N=|\lambda|$, and the
logarithmic negativity is then given by $-\log_2(N)$. The negativity
$N$ is an analytical but lengthy expression of the elements of
\eqref{eq:cov2-tel}, and we will return to special cases in Sec.~V.

In our analysis we shall also encounter the symmetric two-mode
Gaussian state with covariance matrix
\begin{eqnarray} \label{gamma-atom-form}
{\bm \gamma}_{12} = \left(%
\begin{array}{cccc}
  n & 0 & k & 0 \\
  0 & n & 0 & -k \\
  k & 0 & n & 0 \\
  0 & -k & 0 & n \\
\end{array}%
\right).
\end{eqnarray}
Now, $|\lambda_1| = | \lambda_2 | = | n-k|$, $| \lambda_2| =
|\lambda_4|= | n+k|$ and the negativity $N$ reduces to the quantity
$\Delta=\text{Var}(x_1-x_2) = \text{Var}(p_1+p_2)$ known as the EPR
uncertainty~\cite{Giedke03}. We are interested in the case of
entangled symmetric states where
\begin{equation}
\label{EPRuncer} N=\Delta=n-k,
\end{equation}
and $N \in (0; 1]$.

Since the logarithm in the evaluation of $-\log_2 (N)$ is a
monotonic function it is sufficient to consider the argument $N$
$(\Delta)$ in order to quantify the degree of entanglement for
two-mode Gaussian states.

\section{Gaussian covariance formulation}
\label{Sec:Theory} First, as a reference, we study the influence of
loss on the entangled fields of photons leaving an EPR-source
(Fig.~\ref{fig:fig1}(a)). The entanglement in these fields is
maximized when the entangled photons travel equal distances $L/2$ so
the EPR-source is placed in the center of the transmission lines.
For this symmetric setup, the covariance matrix $\gamma_\text{EPR}$
of the two beams is given by Eq.~\eqref{gamma-atom-form} with
initial values of $n$  and $k$ determined by the squeezing parameter
$|\chi |$ for a pure squeezed state, i.e., $n=\cosh |\chi|$,
$k=\sinh |\chi|$, and consequently, the initial EPR uncertainty is
given by $\Delta_0=\exp(-2 |\chi|)$.  In line with our previous
works~\cite{Moelmer04,Madsen04}, we write $\Delta_0=1/r$, and refer
to $r$ as the squeezing parameter. The loss $\epsilon'$ along each
arm changes the canonical $x$ variable according to $x \mapsto
{\sqrt{ 1- \epsilon'}} x+ {\sqrt{\epsilon'}} x_\text{noise}$, where
$x_\text{noise}$ describes the loss-induced vacuum contribution. For
the $p$ quadrature we likewise have $p \mapsto {\sqrt{1- \epsilon'}}
p + {\sqrt \epsilon'} p_\text{noise}$. In terms of the covariance
matrix $\gamma_\text{EPR}$ the mapping is $\gamma_\text{EPR} \mapsto
(1- \epsilon') \gamma_\text{EPR} + \epsilon' I_4$, where $I_4$ is
the $4 \times 4$ identity matrix. Therefore $ n \mapsto
(1-\epsilon') n + \epsilon'$ and $k \mapsto (1-\epsilon') k$, and
the EPR uncertainty, after the transmission, reads
$\Delta_\text{EPR} \mapsto (1-\epsilon') \Delta_0 + \epsilon'$. To
relate to the loss $\epsilon$ in the full transmission line we
assume a uniform loss per propagation distance and hence
$(1-\epsilon') ={\sqrt{1 -\epsilon}}$, and the final result for the
EPR correlation is
\begin{equation}
\label{eq:EPR-delta} \Delta_\text{EPR}= 1 + {\sqrt{1-\epsilon}}
(1/r-1) > 1 - {\sqrt{1 - \epsilon}},
\end{equation}
where the last expression is obtained  in the limit of infinite
squeezing $r\rightarrow \infty$. The smaller the
$\Delta_\text{EPR}$, the higher the degree of entanglement.

We now turn to the case where off-resonant light probes two
separated atomic gases. A continuous-wave light beam which is
linearly polarized along the $x$ direction is sent through samples
of atoms with two degenerate Zeeman levels each described by a Pauli
spin operator ${\bm \sigma}_j$
$(j=1,2,\dots,N_a$)~\cite{Julsgaard01}.  As discussed in detail
elsewhere~\cite{MMbookchapter05}, we describe the interaction and
the measurement process in the time-domain. A continuous-wave light
beam is divided into segments of duration $\tau$ and length $L=c
\tau$ each of which is assumed to be short enough to be accurately
described by a single mode of the electromagnetic field. The
interaction with the atoms and the feedback due to measurement is in
turn described by a succession of interactions with the individual
beam segments. For coherent light and the initially polarized
samples, the uncertainty relations for these variables are
minimized, implying a Gaussian state. The interaction of the samples
with the off-resonant light beam gives rise to a Faraday-rotation
interaction between the macroscopic spin operator ${\bm J}= 1/2
\sum_{j=1}^{N_a} {\bm \sigma}_j$ of the atoms and the Stokes
operator ${\bm S}$ of the light field.

With maximally polarized atomic samples along the $x$ and $-x$
direction, we have the classical relations $J_{x,1} = J_x = -J_{x,2}
= N_a /2$ where the subscripts `1' and `2' refer to the two gases.
We assume a large number of photons in the probe beam such that the
$x$ component of the Stokes vector $S_x$ also can be treated
classically. This leads to the introduction of the following vector
of quantum operators
\begin{eqnarray}
\label{y-vec1} {\bm y} &=& (x_1, p_1, x_2,p_2,x, p)^T\\
\nonumber &=& (\frac{J_{y,1}}{\sqrt{J_x}},
\frac{J_{z,1}}{\sqrt{J_x}}, \frac{- J_{y,2}}{\sqrt{J_x}},
\frac{J_{z,2}}{\sqrt{J_x}}, \frac{S_{y}}{\sqrt{S_x}},
\frac{S_z}{\sqrt{S_x}})^T,
\end{eqnarray}
with canonical commutators for the three independent sets of modes.
 For a probe laser field
propagating in the $y$ direction and being linearly polarized in the
$x$ direction, the Faraday rotation interaction is $\propto
J_{z,i}S_{z}$ ($i=1,2$) and explicitly in terms of the effective
Gaussian variables for a light segment of duration $\tau$ it reads
\begin{equation}\label{noloss}
H_i= \kappa_{\tau,i} p_ip,
\end{equation}
where $\kappa_{\tau,i}$ is the effective coupling constant
$\kappa_{\tau,i} \propto \sqrt{N_a N_{ph}}$ where $N_{ph}$ is the
number of photons in each segment. A polarization rotation
measurement on the optical beam, i.e., a measurement of the $x$
variable of the probe light preserves the Gaussian character of the
states and likewise does the bilinear Hamiltonian of
Eq.~\eqref{noloss}.
\begin{figure}[ttb]
  \includegraphics[width=\columnwidth]{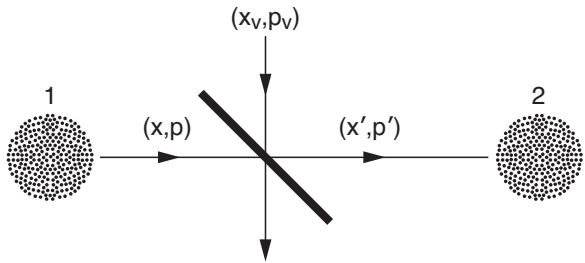}\\
\caption{A beam splitter is used to model the effect of loss of
light intensity $I \rightarrow (1 - \epsilon) I$ through the
transmission line on the $(x,p)$ variables of the field. The output
or detector modes $(x',p')$ are determined by $x'={\sqrt{1
-\epsilon}}x+ {\sqrt \epsilon} x_v$  and $p'={\sqrt{1 -\epsilon}}p+
{\sqrt \epsilon} p_v$ (see text).} \label{fig:fig2}
\end{figure}

In previous work, the effects of loss of atomic spin polarization
and absorption of light within the atomic clouds were
investigated~\cite{Madsen04,Sherson05}. Here we focus on the loss
associated with the light traveling from one gas to the other. This
loss is partly in the transmission line and partly in the
in-coupling devices connecting light to the atomic gas. The loss is
modeled by a reduction ($1-\epsilon$) in the intensity in the probe
beam along the transmission line. We model this loss by a beam
splitter which mixes the incoming fields $x=(a+a^\dagger)/\sqrt{2}$,
$p=(a-a^\dagger)/(i \sqrt{2})$ with vacuum fields $x_v$, $p_v$ (see
Fig.~\ref{fig:fig2}). The beam splitter leads to a mixing of the
operators of the incoming and the vacuum light modes $a \mapsto a'=
\sqrt{1-\epsilon} a + i \sqrt{\epsilon} a_v$, $a^\dagger \mapsto
{a^\dagger}'= \sqrt{1-\epsilon} a^\dagger - i \sqrt{\epsilon}
a_v^{\dagger}$ where a prime denotes the operators of the modes
after the passage of the beam splitter. From these relations which
apply to each polarization component of the light field, and from
the definitions of the Stokes vector components $S_x=a^\dagger_x
a_x/2$, $S_y=(a^\dagger_+a_- -a^\dagger_-a_+)/(2 i)$,
$S_z=(a_+^\dagger a_+ - a^\dagger_-a_-)/2$, the relations between
the input and the primed output modes are easily determined. For the
variables defined in Eq.~\eqref{y-vec1}, we find
\begin{eqnarray}
\label{in-out}
\left(%
\begin{array}{c}
  p \\
  p_v \\
\end{array}%
\right) = \left(%
\begin{array}{cc}
  \sqrt{1-\epsilon} & -\sqrt{\epsilon} \\
  \sqrt{\epsilon} & \sqrt{1-\epsilon} \\
\end{array}%
\right) \left(%
\begin{array}{c}
  p' \\
  p'_v \\
\end{array}%
\right),
\end{eqnarray}
and similarly for the $x$ variables.

The Hamiltonian describing the system when $x$-polarized light
propagating in the $y$ direction is transmitted through the two
gases is $H=\kappa_{\tau,1}p_1p+\kappa_{\tau,2}p_2p'$. We note that
the coupling strength to the second gas is reduced due to the loss
of light intensity along the transmission line, $\kappa_{\tau,2}=
\sqrt{1-\epsilon} \kappa_{\tau,1}$. We denote $\kappa_\tau=
\kappa_{\tau,1}$, and use Eq.~\eqref{in-out} to express the
interaction Hamiltonian in the output modes $(x',p',x_v',p_v')$
\begin{equation}
\label{Hamil2} H=\kappa_\tau \sqrt{1-\epsilon} (p_1 + p_2)p' -
\kappa_\tau \sqrt{\epsilon} p_1 p_v'.
\end{equation}
The atomic $p_1$ and $p_2$ variables are coupled with equal strength
to the output mode $p'$, but the output vacuum noise term couples
exclusively to $p_1$.

The Gaussian variables
\begin{equation}
\label{y-vec2} {\bm y} = (x_1, p_1, x_2,p_2,x'_v,p'_v,x', p')^T,
\end{equation}
describe conveniently the system with transmission loss. The
time-evolution of the operators $y_i$ in ${\bm y}$ is described by
Heisenberg's equations of motion, $i
\partial_t y_i = [ y_i, H]$. The
evolution from $t$ to $t + \tau$ is then determined by the equation
\begin{equation}
\label{evo} {\bm y}(t+\tau) = S {\bm y}(t),
\end{equation}
with $S$ fixed by the Heisenberg equation and given in terms of the
coupling $\kappa_\tau$ and the loss $\epsilon$. Per definition of
the covariance matrix it follows that
\begin{equation}
\label{updategamma} {\bm \gamma}(t+\tau) = S {\bm \gamma}(t) S^T.
\end{equation}
Since the vacuum output components $x_v', p_v'$ describe loss, we
should discard them after the interaction, which simply amounts to
removing the corresponding rows and columns from $\gamma$, which
then becomes a $6 \times 6$ matrix.

We probe the system by measuring the Faraday rotation of the probe
field, i.e., by measuring the light field observable $x'$. The light
field is correlated with the atomic samples and therefore this
measurement will change the atomic state of the system. We denote
the covariance matrix before interaction and detection of light by
\begin{eqnarray}
{\bm \gamma} = \left(%
\begin{array}{cc}
  A & C \\
  C^T & B \\
\end{array}%
\right),
\end{eqnarray}
where the  $4\times 4$ matrix $A$ describes the atomic systems, the
$2 \times 2$ matrix $B$ the output photon field, and the $4 \times
2$ matrix $C$ the correlations between the field part of the system
that is subject to direct measurement and the atomic part that only
feels the measurement-induced back-action. An instantaneous
measurement of $x'$ transforms $A$
as~\cite{GiedkeCirac,Fiurasek02,EisertPlenio,MMbookchapter05}
\begin{equation}
\label{backaction} A \mapsto A' = A - C (\pi B \pi)^-C^T,
\end{equation}
where $\pi =\text{diag}(1,0)$ and where $(\cdot)^-$ denotes the
Moore-Penrose pseudo-inverse. The other parts of the covariance
matrix transform as $B \mapsto {\mathbb I}_{2\times 2}$, $C\mapsto
0_{4 \times 2}$ since a new light segment is used in every
measurement.

In Eq.~\eqref{backaction}, the change in $A$ is proportional to
$\kappa_\tau^2$ which is proportional to the number of photons in
the segment of duration $\tau$, and therefore proportional to
$\tau$. We may then consider the update of the matrix $A$ in the
limit of infinitesimally time steps and obtain a non-linear
differential equation with coupling constant, $\kappa^2 =
\kappa_\tau^2/\tau$.

\section{Asymmetric probing with squeezed light}
\label{sec:asymmetric} Recently, we investigated the output field of
an optical parametric amplifier~\cite{Petersen05b} and formulated a
Gaussian theory in the time-domain that consistently described the
squeezing of the fields for probing times longer than the inverse
bandwidth of squeezing. In an application in precision magnetometry,
we showed that essentially the same degree of squeezing and
precision could be obtained as in the ultra-broadband squeezing case
where the squeezing is just described by the squeezing parameter
$r$.  In the asymmetric case the two gases are subject to the
interaction  \eqref{Hamil2}, and the squeezed light is described by
a covariance matrix $\text{diag}(1/r, r)$, which in the output modes
leads to $T \text{diag}(1,1,1/r, r) T^{-1}$ where $T$ describes the
transformation from $(x_v, p_v, x, p)$ to $(x_v',p_v',x',p')$ (see
Eq.~(\ref{in-out})). The atomic covariance matrix for the two gases
is denoted by $\gamma_{12}$ as in Eq.~(1). We consider the
continuous limit of the update formulas for $\gamma_{12}$ and obtain
a non-linear Ricatti differential equation (see, e.g.,
Refs.~\cite{Moelmer04,Madsen04,Petersen05,Petersen05b,MMbookchapter05}).
The matrices in this equation describe squeezing of $p_1+p_2$, and
noise introduced by the vacuum mode. The time-dependent covariance
matrix which solves the Ricatti equation is on the form of
Eq.~\eqref{eq:cov2-tel} with
\begin{subequations}
\begin{equation}\label{eq:vx1r}
v_{x_1}=
 1+ t \kappa^2(r + 4 \epsilon(1-r)+ 4 \epsilon^2(r-1)),
\end{equation}
\begin{equation}
c_x=t(1-\epsilon)\kappa^2(r + 2 \epsilon(1-r)),
\end{equation}
\begin{equation}
v_{p_1}=\frac{t \kappa^2 r(1-\epsilon) + (1-r)(1-\epsilon)+r} { 2t
\kappa^2 r(1-\epsilon) + (1-r)(1-\epsilon)+r},
\end{equation}
\begin{equation}
c_p=-\frac{t\kappa^2r(1-\epsilon)}{2 t \kappa^2 r (1-\epsilon) +
(1-r)(1-\epsilon) +r},
\end{equation}
\begin{equation}
v_{x_2}=1+t\kappa^2(1-\epsilon)(1+(1-r)(1-\epsilon)),
\end{equation}
\begin{equation}
\label{eq:vp2r} v_{p_2}=\frac{t \kappa^2(1-\epsilon)r +
(1-r)(1-\epsilon) + r}{2 t\kappa^2 (1-\epsilon) r +
(1-r)(1-\epsilon)+r}.
\end{equation}
\end{subequations}
We use these expressions in Sec.~\ref{sec:quantifying} to quantify
the degree of entanglement between the two gases. We shall see that
although some of these expressions diverge individually for large
$t$, the logarithmic negativity approaches a constant value.

\section{Symmetric probing with squeezed light}
\label{sec:symmetric}
We now turn to the symmetric case where equal
amounts of noise is added to all four quadratures. The symmetric
setup is obtained by sending light of different directions and
polarizations through the two atomic samples, i.e., the gases are
subject to the following sequence of effective interactions $H_1 =
\kappa _\tau \sqrt{ 1 - \epsilon} (p_1 + p_ 2) p_1' - {\sqrt
\epsilon} p_{v1}' \kappa_\tau p_1$, $H_2 = \kappa_\tau \sqrt{1 -
\epsilon} (p_1 + p_ 2) p_2' - {\sqrt \epsilon} p_{v2}' \kappa_\tau
p_2$, $H_3 = \kappa_\tau \sqrt{1 - \epsilon} (x_1 - x_ 2) x_3' -
{\sqrt \epsilon} x_{v3}' \kappa_\tau x_1$, and $H_4 = \kappa_\tau
\sqrt{1 - \epsilon} (x_1 - x_ 2) x_4' + {\sqrt \epsilon} x_{v4}'
\kappa_\tau x_2$, where $p_{iv}'$ and $x_{iv}'$ are different vacuum
modes. Each of these Hamiltonians leads to a propagation matrix $S$
via Heisenberg's equations of motion. Tracing over all other degrees
of freedom than those of the two atomic gases and adding the
differential equations obtained for the four Hamiltonians above
leads to a single Ricatti equation and if we probe with equal
strengths and equal squeezing parameter (but $r \rightarrow 1/r$
when $p'$ components are measured instead of $x'$), the
 covariance matrix is of the symmetric form
\eqref{gamma-atom-form}. The quantity $N=\Delta = n-k$ quantifies
the degree of entanglement, and solves the equation
\begin{equation}
\label{eq:Delta-evo}
\partial_t \Delta = a - b \Delta^2,
\end{equation}
with growth rate for the noise
\begin{equation}
\label{eq:noise-a} a=\kappa^2 \epsilon (1-\epsilon + r\epsilon),
\end{equation}
while the term driving the entanglement formation is given by
\begin{equation}
\label{eq:squeezing-b} b= \frac{4 \kappa^2 r
(1-\epsilon)}{1-\epsilon + r \epsilon}.
\end{equation}
The solution to Eq.~\eqref{eq:Delta-evo} with initial condition
$\Delta(t=0)=1$ is found by quadrature and is given by
\begin{equation}
\label{eq:Delta-res} \Delta(t)= \frac{\cosh({\sqrt{ab}} t) +{\sqrt
\frac{a}{b}} \sinh({\sqrt{ab}} t) } {\cosh({\sqrt{ ab}} t) +{\sqrt
\frac{b}{a}} \sinh({\sqrt{ab}} t) }.
\end{equation}

In the presence of transmission line loss and squeezed light it
follows that the steady-state value for the EPR uncertainty is
\begin{equation}
\label{eq:ss-r} \Delta = \frac{1-\epsilon+r\epsilon}{2}
\sqrt{\frac{\epsilon}{r(1-\epsilon)}}.
\end{equation}
Since $\Delta$ must be less than unity for the two gases to be
entangled, Eq.~\eqref{eq:ss-r} shows that for unsqueezed probe light
($r=1$) it is possible to entangle the two gases only if $\epsilon <
4/5$.

The minimum value of $\Delta$ (maximal entanglement) is
\begin{equation}
\label{eq:minimumEPR} \Delta_\text{min}=\epsilon.
\end{equation}
and is obtained for the optimal squeezing
\begin{equation}
\label{eq:optimal-r} r_\text{opt} = \frac{1-\epsilon}{\epsilon}.
\end{equation}
This equation has as a consequence a property that at first sight
seems surprising: for $\epsilon < 1/2$ the minimum in $\Delta$
occurs for squeezed light ($r > 1$) as expected,  but for $\epsilon
> 1/2$ the minimum in $\Delta$ is obtained for antisqueezed light, $r < 1$.
This happens because the squeezed light input modes contribute to
the undetected output mode which acts as a noise term
$-{\sqrt{\epsilon}} p_{v}'\kappa_\tau p_1$ on the first gas in
Eq.~\eqref{Hamil2}.  By Heisenberg's equation of motion for $x_1$,
we obtain from this part of the Hamiltonian the mapping $x_1 \mapsto
x_1' =x_1 -{\sqrt{\epsilon}}p_{v}'\kappa_\tau$ which means that the
added noise is determined by the variance $\epsilon \kappa^2
\text{Var}(p_{v}')= 1/2 \kappa^2 \epsilon( 1 -\epsilon + r \epsilon)
$ which is precisely half the factor entering the noise term in
Eq.~\eqref{eq:noise-a}. When we probe with light that is squeezed
(antisqueezed) in the $x$ quadrature, the $p$ quadrature is
antisqueezed (squeezed) and the noise term is sensitive to the noise
in this antisqueezed (squeezed) component

\section{Quantifying the entanglement of the channel}
\label{sec:quantifying}
\begin{figure}[ttb]
  \includegraphics[width=\columnwidth]{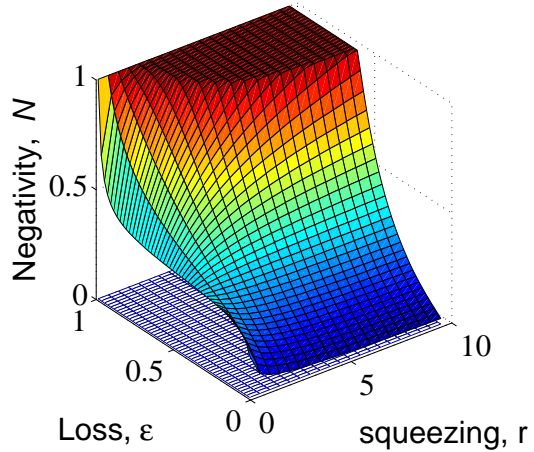}\\
\caption{(Color online). Negativity at asymptotic times as a
function of loss, $\epsilon$, and squeezing parameter $r
\in[0.1;10]$ for the atomic asymmetric case where the light is only
sent one way through the two samples, and where only the $x'$
quadrature is probed.
}\label{fig:fig3}
\end{figure}
\begin{figure}[ttb]
  \includegraphics[width=\columnwidth]{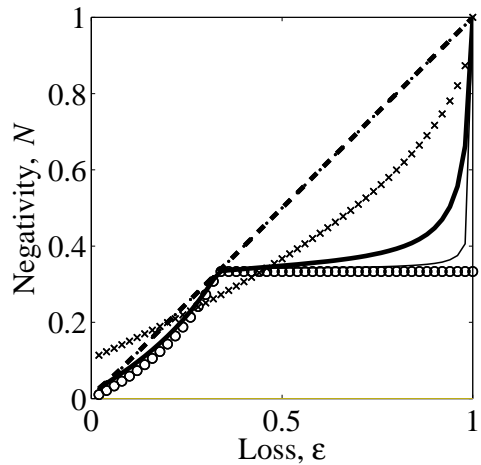}\\
\caption{Optimal vales of the negativity for $r \in [0.1;10]$ at
asymptotic times as a function of loss $\epsilon$ through the
transmission line. The full curve corresponds to the case where the
light is only sent one way through the two gases, and where only the
$x'$ quadrature is probed. The thin full curve corresponds to the
case $r_\text{min} =0.01$.
The long-dashed curve is for the symmetric case with $N=\epsilon$.
The crossed curve is for the case of light emitted from an optical
parametric oscillator \eqref{eq:EPR-delta} and the circles denote
the asymptotic form of the full curve in the long-time and
infinite-squeezing limit $r \gg 1$, $r \ll 1$,
\eqref{Nlimit-small-epsi}, \eqref{eq:constantN}.}\label{fig:fig4}
\end{figure}

For the asymmetric setup with light sent only one way through the
samples,  Fig.~\ref{fig:fig3} shows a surface plot of the negativity
$N$ for long probing times as a function of $\epsilon$ and $r$ for
$r\in [0.1; 10]$. The negativity $N$ is given by an analytical but
very lengthy expression composed of sums of products of the
variances and covariances from \eqref{eq:vx1r}-\eqref{eq:vp2r}.

The results are summarized and compared with the symmetric
interaction case and the EPR fields in Fig.~\ref{fig:fig4}. For a
given loss, the full curve shows the optimum negativity from
Fig.~\ref{fig:fig3} assuming physical limits to the available
squeezing $r_\text{min} = 0.1$, $r_\text{max}=10$. The thin full
curve shows the corresponding result for $r_\text{min}=0.01$, and
the open circles show the results obtained with infinite squeezing:
For losses less than $1/3$ and in the limit of infinite squeezing
($r \gg 1$) we obtain
\begin{equation}
\label{Nlimit-small-epsi} N=\frac{\epsilon}{2-3\epsilon}.
\end{equation}
For arbitrarily high losses, we find surprisingly that the
negativity  approaches a constant for antisqueezed light ($r \ll 1$)
\begin{equation}
\label{eq:constantN} N= 1/3,
\end{equation}
and we find that at the particular loss of $\epsilon = 1/3$, $ N =
1/3$ for all values of the squeezing parameter, $r$. These results
mean that no matter the loss it is always possible (assuming that
the appropriate squeezed probe light source is available) to
entangle the two gases to a  high level.

The crossed curve is given by Eq.~\eqref{eq:EPR-delta} for  the
symmetric EPR light source operated at a finite degree of squeezing
$r=r_\text{max}=10$.  The long-dashed curve describes the minimum
value of $N$ for the symmetric atomic case (maximum entanglement)
$N=\Delta_\text{min} =\epsilon$ obtained by the optimal squeezing of
Eq.~\eqref{eq:optimal-r}.

From Fig.~\ref{fig:fig4} we see that for small loss and finite
squeezing the light-atom system leads to higher entanglement than
achievable with the EPR light source for the same degree of
squeezing. For the asymmetric atom-light case we see from the full
curve in Fig.~\ref{fig:fig4} that the atom-light setup leads to a
higher degree of entanglement than the EPR light source except in a
narrow range around $\epsilon = 1/3$. In particular for losses
exceeding $\sim$ 40 \%, the asymmetric setup presents an efficient
way to generate a high and constant degree of entanglement.

\section{Discussion}
\label{sec:Tele} In summary, we have studied the effects of (light)
losses on the achievable entanglement between two locations coupled
by a lossy transmission line. We characterize the entanglement by
the argument  $N$ of the logarithmic negativity $-\log_2(N)$.  The
entanglement of atomic samples by optical probing is experimentally
interesting because it works with classical optical sources and
because the entangled states are stored in stationary media for
potential later use. If only coherent light is available, the
entanglement obtained with light being sent in both directions
between the gases and with probing of suitable combinations of both
the $x$ and $p$ atomic variables yields
$N=\sqrt{\epsilon/(4-4\epsilon)}$, and the gases become entangled as
long as the losses are below 80 \%. When squeezed light is
available, we both have the possibility to simply prepare and send
entangled light pulses over the transmission channel and we may use
the squeezed light to improve the atomic probing schemes. For low
losses, and with no limits to the degree of squeezing, the mere
emission of entangled light pulses yields the largest entanglement,
but if the degree of available optical squeezing is limited, it
becomes advantageous to use probed atomic samples, and, quite
surprisingly we showed that for large losses, a one-way probing
scheme with anti-squeezed light leads to a finite negativity $N=1/3$
and a logarithmic negativity as large as $\log_2(3)$ for any value
of the loss. When the loss approaches unity, the required degree of
anti-squeezing and probing time get larger, but as shown in Fig.~4.,
realistic squeezing levels suffice to give surprisingly small $N$
for large losses.

Entangled states have been proposed for various quantum
communication tasks, and since losses set a practical limit on the
distance over which these tasks can be carried out, a scheme that
attains high entanglement despite losses is of course interesting.
To investigate if our entanglement is useful, we present a brief
analysis of the achievements of the entangled gases for continuous
variables teleportation~\cite{Braunstein98}. We have recently
presented a general analysis of fidelities for Gaussian state
transformations, including teleportation ~\cite{MM-fidelity05}, and
for an entangled channel of the form (3), we reproduced the known
fidelity for teleportation of an unknown coherent state
\begin{equation} F=\frac {1}{ 1+n-k }=\frac{1}{1+\Delta}.
\end{equation}
This result applies to the case of symmetric probing, and with
optimum use of squeezed probing light \eqref{eq:minimumEPR} the
fidelity gets as high as
\begin{equation} \label{eq:Fopt} F^\text{opt}=\frac{1}{1+\epsilon},
\end{equation}
approaching the classical limit of $F_\text{classical}=
1/2$~\cite{Hammerer05} when the whole light field is lost.

The entangled state obtained from the asymmetric probing can also be
applied in the Braunstein-Kimble teleportation protocol, in which
case the fidelity is expressed in terms of the variances of the
non-local variables of the entangled channel
$p_+=(p_1+p_2)/{\sqrt{2}}$ and $x_-=(x_1-x_2)/{\sqrt{2}}$:
\begin{equation}
\label{F-channel1}
F=\frac{1}{\sqrt{(1+2\text{Var}(p^{(12)}_+))(1+2\text{Var}(x^{(12)}_-))}}.
\end{equation}
The theory in Sec.~\ref{sec:asymmetric} provides the values
$\text{Var}(p^{(12)}_+)= 1/(2(1+ \beta t))$, $\text{Var}(x_-^{(12)}
= (1+ \alpha t)/2$ with
$\beta=2(1-\epsilon)r\kappa^2/(1+r\epsilon-\epsilon)$ and
$\alpha=\kappa^2 \epsilon (1+r \epsilon -\epsilon)/2$, and instead
of applying the protocol directly we suggest to locally anti-squeeze
the $p$'s, and hence $p^{(12)}_+$, and squeeze the $x$'s, and hence
$x^{(12)}_-$, so that $\text{Var}(p^{(12)}_+) \rightarrow s
\text{Var}(p^{(12)}_+)$, and $\text{Var}(x^{(12)}_-) \rightarrow
\text{Var}(x^{(12)}_-)/s$. The maximum value of (\ref{F-channel1})
is obtained for $s=
\sqrt{\text{Var}(x^{(12)}_-)/\text{Var}(p^{(12)}_+)}$,
\begin{equation}
F= \frac{1}{1+2\sqrt{\frac{1+\alpha t}{1+\beta t}}} \rightarrow
\frac{1}{1+2\sqrt{\alpha/\beta}}\ \ (t \rightarrow \infty).
\end{equation}
This expression is always less than or equal to $1/(1+\epsilon)$,
the optimum in the case of symmetric probing \eqref{eq:Fopt} with
squeezed light, and equality is obtained for probing with
squeezed/anti-squeezed light with a finite squeezing factor,
$r=(1-\epsilon)/\epsilon$, which is incidentally also the optimal
optical squeezing in the case of symmetric probing. For large
$\epsilon$, e.g., $90 \%$, this value of $r$ does indeed lead to
significant entanglement, cf. Fig.4, but the teleportation fidelity
behaves "normally" and approaches the classical value of $1/2$ when
$\epsilon \rightarrow 1$. The surprisingly high entanglement does
not provide a scheme for high fidelity teleportation!

We note that we have not really proven that we made optimum use of
the entangled state by the straightforward use of the
Braunstein-Kimble protocol, but we can give an independent argument
for why (\ref{eq:Fopt}) must be the optimum teleportation fidelity
in the case of the asymmetric probing scheme: Imagine a modified
physical setup where the transmission line and the second gas of our
analysis is replaced by a loss less beam splitter arrangement and
$M$ separate gases, which each receive a fraction $1/M$ of the field
after transmission through the first gas. Any one of these gases
will become entangled with the first gas as described by the above
theory, if we identify the transmitted fraction $1-\epsilon$ with
$1/M$. The teleportation of an unknown coherent state is
accomplished by performing a joint measurement on the incident state
and the first gas, and communicating the outcome to the other site,
where a local operation on the quantum system establishes the output
state. Since we share the entangled state with $M$ different sites,
we are thus able to teleport the same state to $M$ different sites,
i.e., produce $M$ approximate clones of the incoming state. The
fidelity for each of these clones, however, has been shown by a
different argument to be limited by the value
$M/(2M-1)=1/(1+\epsilon)$~\cite{Cerf00}, which therefore must also
be the upper limit of our teleportation fidelity.

In conclusion, without violating fundamental results on the
achievements of distillation, purification and cloning of Gaussian
states, we have identified a protocol, that leads to finite
entanglement between systems which are coupled by even very lossy
transmission lines. If the probing field is split and distributed
evenly between a number of partners, our analysis identifies instead
a new kind of "entanglement polygamy", where a single system can
share finite entanglement with an unlimited number of other systems.
Unlike a recent proposals for multi-partite Gaussian GHZ-type
states~\cite{Adesso05}, where the entanglement can be concentrated
and used for high-fidelity teleportation between any pair by
measurements on all the other systems, our entanglement is already
sizable, but not correspondingly useful for teleportation. The
logarithmic negativity has been criticized for not always giving a
reliable characterization of the entanglement of quantum systems. In
our case, however, we have observed that also the more involved
Gaussian entanglement of formation~\cite{wolf04} is finite for large
losses. Rather than  demonstrating a weakness of specific
entanglement measures, we believe that the present study is a
contribution to our discovery of different varieties of useful and,
possibly, useless entanglement.

\begin{acknowledgments}
We thank J. Sherson and U. V. Poulsen for useful discussion. L.B.M.
was supported by the Danish Natural Science Research Council (Grant
No. 21-03-0163) and the Danish Research Agency (Grant. No.
2117-05-0081).
\end{acknowledgments}


\end{document}